# The ferroelectric – paraelectric transition in $BiFeO_3$: Crystal structure of the orthorhombic β-phase

Donna C. Arnold[1], Kevin S. Knight[2], Finlay D. Morrison[1], Philip Lightfoot[1*]

[1] *School of Chemistry, University of St Andrews, North Haugh, St Andrews, Fife, KY16 9ST, UK*

[2] *ISIS Facility, Rutherford Appleton Laboratory, Chilton, Didcot, OX11 0QX, UK*

[*] *To whom correspondence should be addressed: e-mail: pl@st-andrews.ac.uk, Tel: +44-1334463841, Fax: +44-1334463805.*

## Abstract

A detailed investigation using variable temperature powder neutron diffraction demonstrates that $BiFeO_3$ undergoes a phase transition from the ferroelectric α phase (rhombohedral, R3c) to a paraelectric β phase (orthorhombic, Pbnm) between 820 °C and 830 °C. Co-existence of both phases over a finite temperature interval, together with abrupt changes in key structural parameters, confirm that the transition is first-order. The β phase corresponds to the $GdFeO_3$-type perovskite. The metastability of $BiFeO_3$ relative to $Bi_2Fe_4O_9$ above 810 °C precludes observation of a reported cubic γ phase above 925 °C under the present experimental conditions.

## Introduction

Multiferroic materials have attracted a lot of attention primarily due to their potential applications including data storage, spintronics and microelectronic devices.[1,2] By far the most widely studied multiferroic material is $BiFeO_3$, primarily since its electrical and magnetic ordering both occur above room temperature: ferroelectric $T_C$ ~ 810 – 830 °C[3,4] and antiferromagnetic $T_N$ ~ 350 – 370 °C.[5] Despite extensive studies on bulk $BiFeO_3$ there are still many contradictions in the literature. This is partly due to problems in preparing high quality, single phase $BiFeO_3$ and authors have presented methods ranging from rapid liquid-phase sintering[6] to the use of ultra pure $Bi_2O_3$[7] in order to overcome these issues. In particular there has been no conclusive data on the evolution of the crystal structure as a function of temperature above 650 °C and the manner in which the structure changes at the reported ferroelectric transition temperature $T_C$, and above. In early work as many as eight anomalies were found in physical properties as a function of temperature[8], although only three were suggested on the basis of electrical and magnetic measurements.[9] Of these, the upper two coincide with the transitions at $T_N$ and $T_C$, where the third was reported at a lower temperature of approximately 190 °C. Variable temperature X-ray powder diffraction (XRPD) studies[10], suggested two anomalies in lattice parameters at $T_N$ and $T_C$ and this was supported by the powder neutron diffraction (PND) study of Fischer *et al.*[11] who found no definitive evidence of a phase transition at 190 °C. These authors reported that the rhombohedral, R3c, room temperature structure is maintained up to 605 °C, although with a gradual reduction in the octahedral tilt within the rhombohedral phase in this temperature regime. The highest temperature reported here is, however, still significantly below $T_C$. More recently, this study was extended by Palewicz *et al.*[12], also using PND, who were able to track precise trends in various structural parameters, including a decrease in octahedral rotation, a decrease in $Bi^{3+}$ cation displacement and an increase in the Fe-O-Fe angles, up to a temperature of 650 °C. These authors reported a significant decomposition to $Bi_2Fe_4O_9$ at 700 °C; a problem which occurs in many studies of $BiFeO_3$ at elevated temperatures.

The most complete study of the phase diagram of $BiFeO_3$[13] based on thermal analysis, spectroscopic, diffraction and other methods, suggests three distinct solid phases above room temperature and below the melting point (~960 °C): the rhombohedral α-phase, below $T_C$, an intermediate β-phase, in the region 830 – 925 °C, and a cubic γ-phase in the region 925 – 933 °C before decomposition and subsequent melting. In this paper, evidence is given for an orthorhombic (pseudo-tetragonal) β-phase, based on both Raman and polarised light measurements, and a cubic γ-phase[13] in disagreement with other recent work suggesting a cubic β-phase[14] based solely on Raman data. Palai *et al.* report the appearance of the cubic γ-phase to coincide with a metal-insulator transition. More recently Kornev *et al.*[15] applied first-principles techniques to the high temperature properties of bulk $BiFeO_3$. These calculations predicted that above $T_C$ the structure adopts a tetragonal phase (space group I4/mcm) which is associated with antiferrodistortive motions, before transforming to a cubic phase (Pm-3m) at a temperature of approximately 1167 °C, much higher than that observed by Palai *et al.* In parallel with the simulations the authors conducted a high temperature XRPD study, which suggested that above $T_C$ the observed peaks could only be indexed satisfactorily with a monoclinic (but pseudo-tetragonal) unit cell with dimensions a ~ 5.61, b ~ 7.97, c ~ 5.65 Å, and space group C2/m. A follow-up paper gave a fuller model for this phase, and suggested the true space group may be $P2_1/m$.[16] However, due to decomposition of the $BiFeO_3$ at ~ 925 °C, with subsequent melting at ~ 965 °C, they were unable to observe the cubic γ-phase predicted in their calculations and observed by Palai *et al.*[13,15] In another recent high temperature XRPD study, Selbach *et al.* also observed a first-order phase change at $T_C$, but assigned the paraelectric β-phase directly above $T_C$ to a centrosymmetric rhombohedral symmetry, space group R-3c.[17]

In this letter we present the most detailed crystallographic study yet of the nature of the ferroelectric – paraelectric transition in $BiFeO_3$, using PND, which shows unambiguously that the paraelectric β-phase adopts orthorhombic symmetry. We present the first fully refined crystallographic model for this elusive phase, which is fundamentally distinct from all the previous models.

Single phase $BiFeO_3$ was prepared using a similar method to that originally reported by Achenbach[18] and later adopted by Zhou *et al.*[19] Stoichiometric ratios of $Fe_2O_3$ (Aldrich, ≥ 99 %) and $Bi_2O_3$ (Aldrich, 99.9 %) were reacted with a 6 mol % excess of $Bi_2O_3$ at 800 °C for 5 hours. After calcination the material was ground into a fine powder and leached with 2.5 M nitric acid under continuous stirring for 2 hours and washed thoroughly with distilled water. The resulting powder was then dried at 400 °C for 1 hour. X-ray diffraction patterns collected for the material obtained post-leaching confirmed single phase $BiFeO_3$, according to XRPD.

PND data were collected on the high resolution powder diffractometer (HRPD) at ISIS, over a temperature range of 100 °C to 900 °C. Short data collections (~20 minutes) were undertaken at 760, 770, 780, 790, 805, 810, 815, 835, 840, 850, 870, 880 and 900 °C, and 1 hour datasets at 100, 300, 500, 750, 800, 820, 825, 830, 860 and 890 °C. HRPD offers the advantage of a variable incident wavelength with fixed detector geometry. Prior to this experiment the sample was dried and sealed within a quartz ampule, which was contained within a cylindrical vanadium can.

Crystallographic analysis of the PND data was performed by the Rietveld method using the GSAS suite.[20] The refinements included individual isotropic atomic displacement parameters for all atoms of the majority phase, together with the usual profile coefficients. Full details of the refinements have been deposited with ICSD.[21]

At temperatures between 100 and 815 $^{o}$C the crystal structure can be described in the polar space group R3c ($a^-a^-a^-$ Glazer tilt system) consistent with all previous reports.[12,13,15,16,17] Below $T_N$ magnetic Bragg peaks associated with the long-range magnetic ordering of the $Fe^{3+}$ magnetic moments can be observed. We have made no further study of the magnetic ordering, which has been discussed at length elsewhere.[22-24] Moreover, below 810 $^{o}$C there is no evidence of unindexed peaks, confirming that $BiFeO_3$ is indeed single phase. In the 810 $^{o}$C and 815 $^{o}$C datasets there is evidence that $BiFeO_3$ begins to very slowly decompose into $Bi_2Fe_4O_9$, consistent with the reports of Palai *et al.*[13] At these temperatures, and higher, $Bi_2Fe_4O_9$ was included as a secondary phase in the refinements, using the structural model proposed by Tutov *et al.*[25]: at 815 $^{o}$C approximately 2.5 % of this impurity phase is observed.

As the α-phase evolves with increasing temperature, changes in the lattice parameters and unit cell volume (Fig. 1(a)) follow the same trends observed by previous authors.[12,13,16] The oxygen octahedral rotation, ω, and the 'polar' shifts of the $Bi^{3+}$ (tc) and $Fe^{3+}$ (sc) ions from their 'ideal' positions (c is the lattice parameter; t and s parameters are defined by Megaw *et al.*[26] and Moreau *et al.*[27] and summarised by Fischer *et al.*[11]) are plotted in Fig. 1. Palewicz *et al.* reported a decrease in oxygen octahedral rotation over the temperature range they studied.[12] We confirm that this trend continues as $T_C$ is approached. Both the $Bi^{3+}$ and $Fe^{3+}$ ion polar shifts decrease with increasing temperature. The former is consistent with the data presented by Palewicz *et al.*[12] However, these authors failed to observe the $Fe^{3+}$ ion shift over the temperature range they studied. Our data clearly show that whilst the $Fe^{3+}$ ion shift from its ideal perovskite position is reasonably constant in the temperature regime studied by the previous authors it also undergoes a sharp decrease at temperatures > 700 $^{o}$C as it approaches the phase change to orthorhombic symmetry at $T_C$.

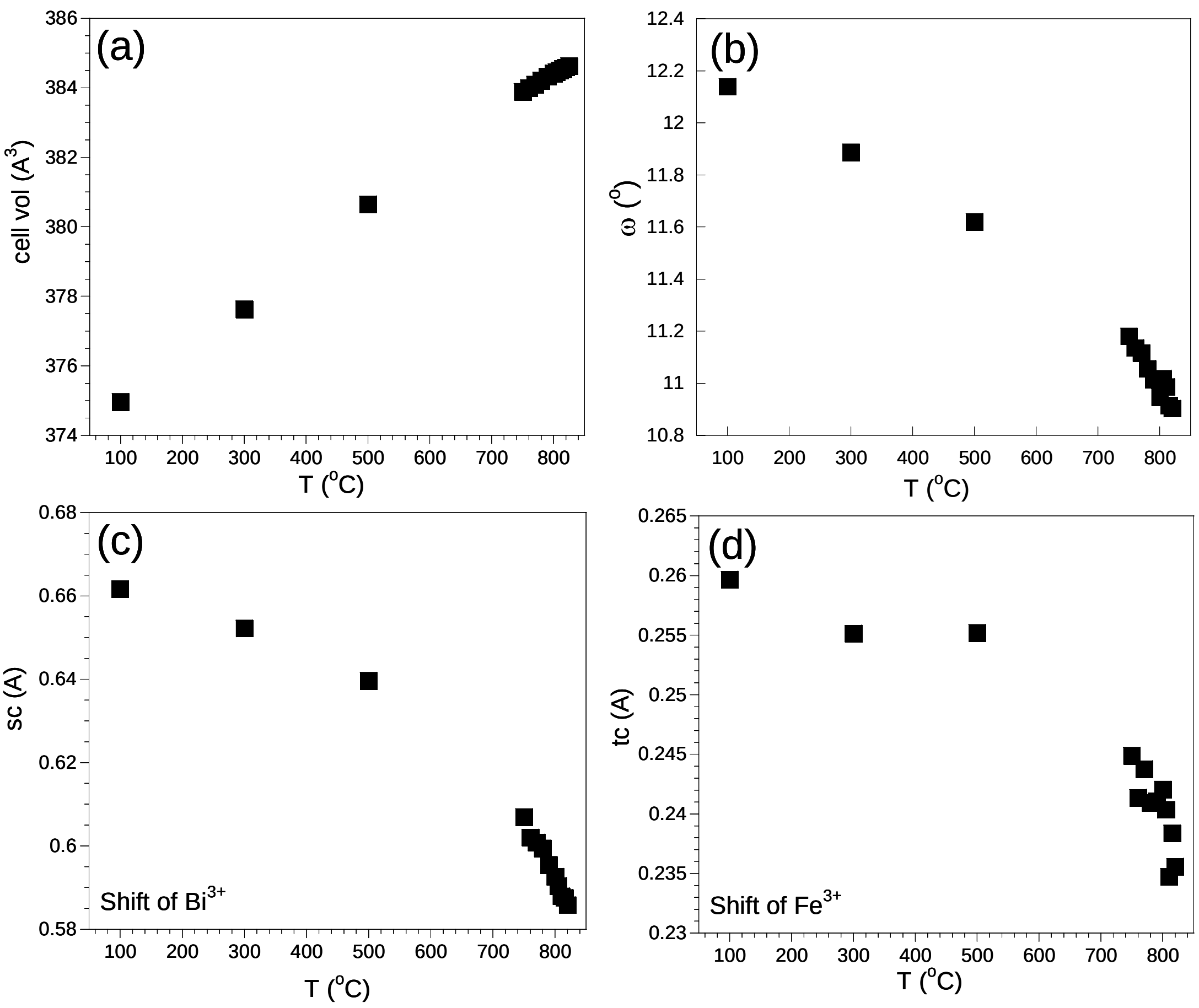

**Figure 1:** Thermal evolution of the R3c α-phase (a) Cell volume, (b) Oxygen octahedral rotation angle, ω, (c) Polar shift of $Bi^{3+}$ ion from its ideal perovskite position, sc, and (d) Polar shift of $Fe^{3+}$ ion from its ideal perovskite position, tc. Error bars are smaller than the symbols.

Through the temperature range 820 $^{o}$C – 830 $^{o}$C $BiFeO_3$ clearly undergoes a dramatic first order phase transition, where the co-existence of both perovskite phases can be seen at 825 $^{o}$C, with transformation complete by 830 $^{o}$C. In addition, in all refinements the phase percentage contribution from the $Bi_2Fe_4O_9$ secondary phase continues to grow slowly consistent with the slow decomposition of $BiFeO_3$. At 820 $^{o}$C, Rietveld shows a phase mixture of ~ 93% α, 3% β and 3% $Bi_2Fe_4O_9$; at 825 $^{o}$C the phase fractions are 12 %, 82% and 6%, respectively. The β-phase of $BiFeO_3$ persists up to a temperature of approximately 890 $^{o}$C before complete decomposition into $Bi_2Fe_4O_9$. Although our observed temperature for the α to β transition is consistent with previous work[13,15-17] our conclusions regarding the nature of this transition, in particular the structure of the β-phase, are markedly different. Close inspection of the PND data clearly shows a step-shift in peak position for the peak at d ~ 2 Å (corresponding to the 200 cubic perovskite reflection), coupled with a distinct splitting (Fig. 2) confirming the change of symmetry and suggesting a transformation to the orthorhombic system. The rhombohedral R-3c model for the β-phase proposed by Selbach[17] can therefore be immediately excluded. Further inspection of the 830 $^{o}$C data revealed that all the observed reflections not accounted for by the $Bi_2Fe_4O_9$ impurity (ie. due to the β-$BiFeO_3$ phase) could be indexed in a metrically orthorhombic unit cell approximately 5.61 x 5.65 x 7.97 Å. Although this is metrically consistent with the monoclinic cell suggested by Kornev[15,16] the presence of key reflections violating C-centring, the absence of any peak splitting consistent with further symmetry lowering, and the observation of systematic absences demonstrating two glide planes, led us to postulate the orthorhombic space group Pbnm. Our model for the β-phase was therefore that of $GdFeO_3$,[28] which exhibits the most common tilting distortion of a perovskite, corresponding to the Glazer tilt system $a^-a^-b^+$. The refinement profile and the derived atom parameters are given in Figure 3 and Table 1, respectively.

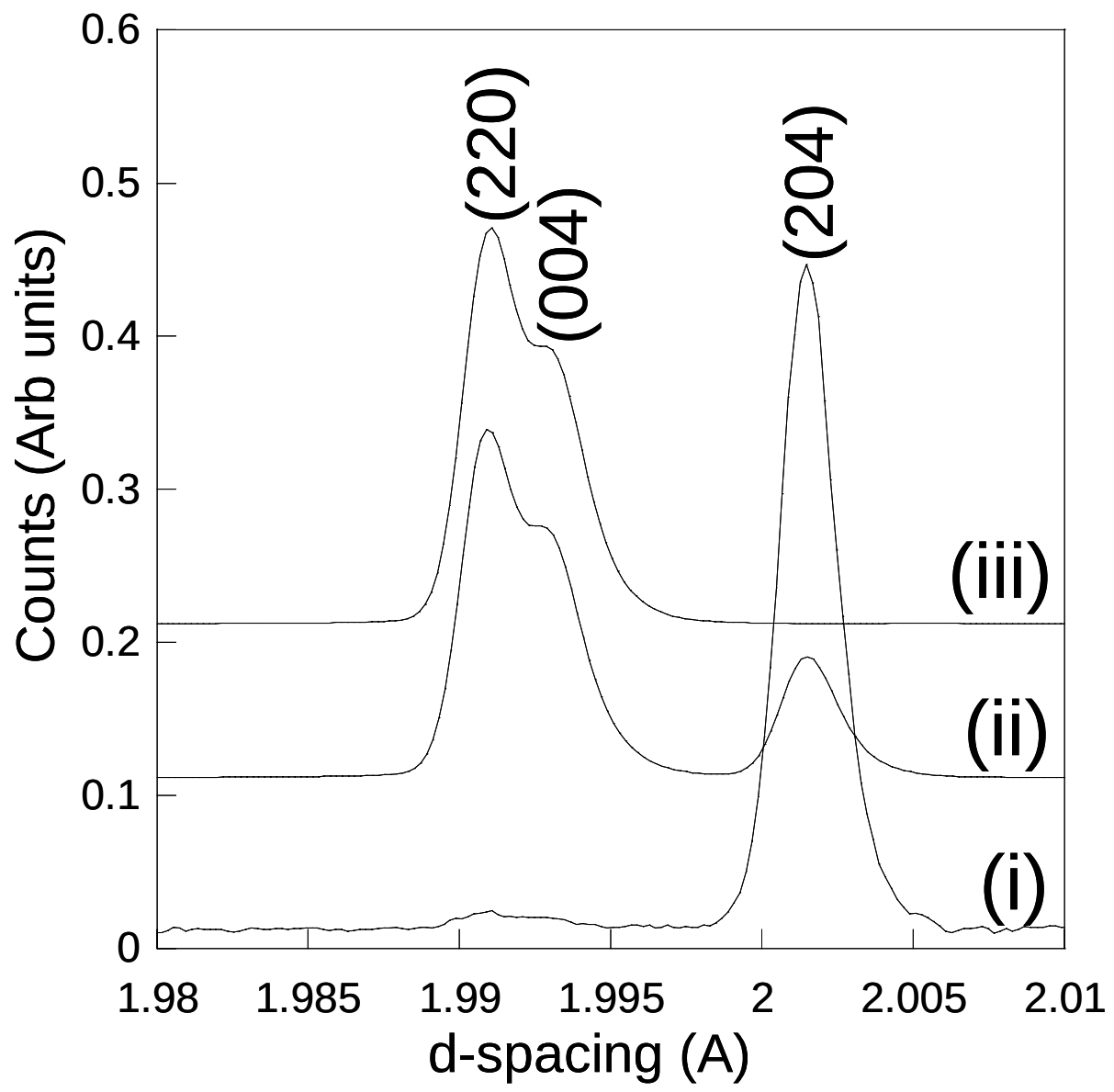

**Figure 2:** Comparison of the pseudo-cubic perovskite 200 peak passing through the α−β transition between rhombohedral, R3c, and orthorhombic, Pbnm, symmetry. (i) 820 $^{o}$C, (ii) 825 $^{o}$C and (iii) 830 $^{o}$C. Indices correspond to rhombohedral (hexagonal setting) and orthorhombic unit cells for (i) and (iii), respectively.

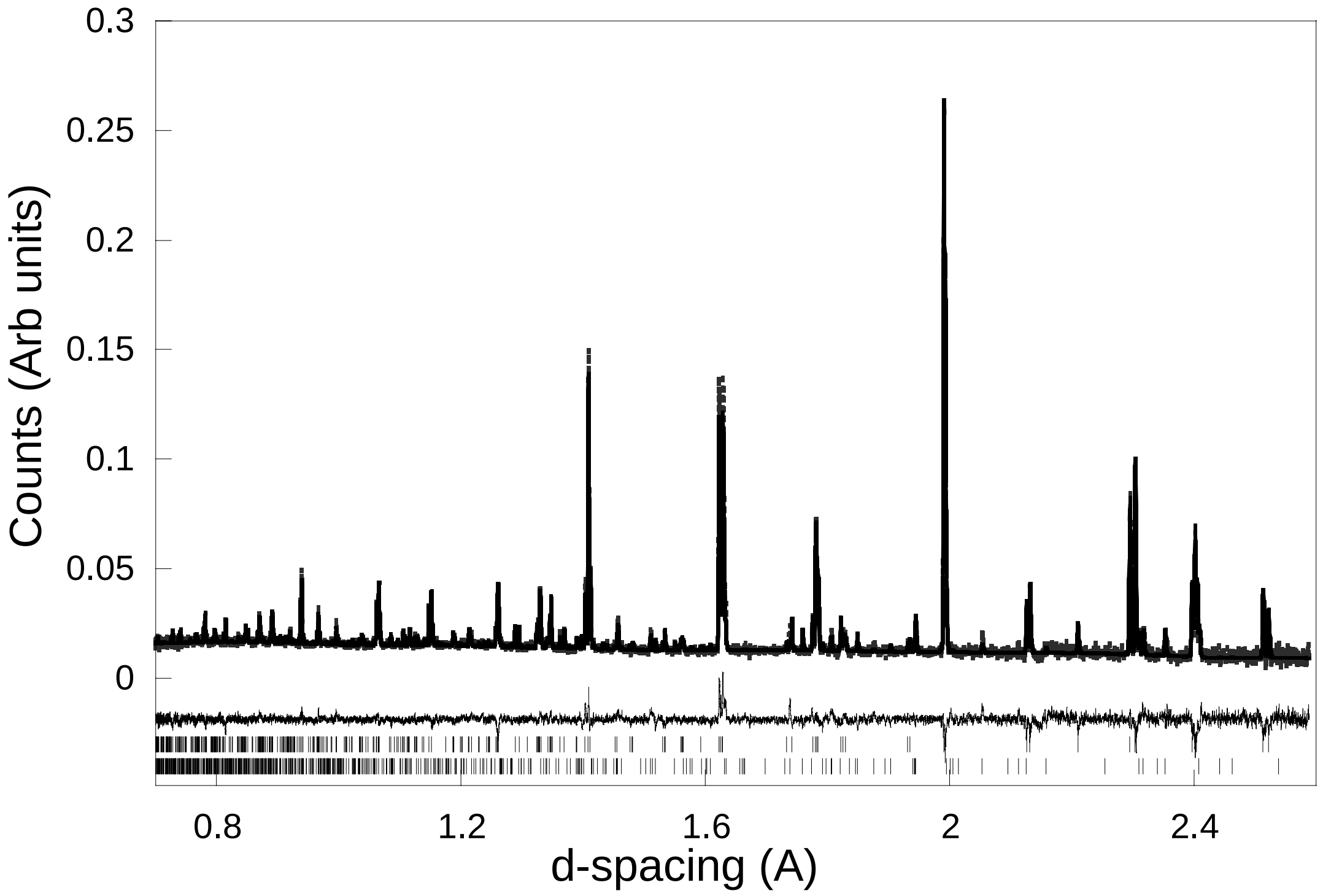

**Figure 3:** Refinement profile at 830 °C. The majority contribution is the β-$BiFeO_3$ Pbnm phase, represented by the upper tick marks, and the $Bi_2Fe_4O_9$ impurity phase is represented by the lower tick marks.

**Table 1(a)** Structural parameters for β-$BiFeO_3$ from the 830 °C data set*. Space group Pbnm, a = 5.61328(5) Å, b = 5.64697(5) Å, c = 7.97114(7) Å, V = 252.669(4) Å$^3$. $\chi^2$ = 1.963, wRp = 8.03 %, Rp = 7.85 %.

| Atom | Site | x | y | z | U(iso) x 100 / Å$^2$ |
|---|---|---|---|---|---|
| Bi | 4c | 0.9995(5) | 0.0233(6) | 0.25 | 6.30(5) |
| Fe | 4a | 0.5 | 0 | 0 | 1.23(3) |
| O1 | 4c | 0.0664(4) | 0.4799(6) | 0.25 | 3.96(8) |
| O2 | 8d | 0.7118(3) | 0.2893(3) | 0.0325(2) | 4.13(6) |

*Refined phase fractions: β-$BiFeO_3$ 95.5%, $Bi_2Fe_4O_9$ 4.5%

**Table 1(b)** Selected bond lengths (Å) and angles (deg) for β-$BiFeO_3$ at 830 $^oC$

| | | | |
|---|---|---|---|
| Fe-O1 x 2 | 2.025(2) | Bi-O1 | 2.605(4) |
| Fe-O2 x 2 | 2.031(4) | Bi-O1 | 2.449(3) |
| Fe-O2 x 2 | 2.037(2) | Bi-O2 x 2 | 2.805(3) |
| | | Bi-O2 x 2 | 2.482(3) |
| | | Bi-O2 x 2 | 2.759(3) |
| Fe-O1-Fe | 157.9(1) | | |
| Fe-O2-Fe | 157.1(1) | | |
| Bi-Fe | 3.349(2) | | |

With increasing temperature $BiFeO_3$ gradually decomposes to $Bi_2Fe_4O_9$; the phase fraction of $Bi_2Fe_4O_9$ increases from about 7% at 835 $^oC$ to about 64% at 870 $^oC$. Nevertheless, the evolution of the $BiFeO_3$ structure can be followed quantitatively. The lattice parameters of the β-$BiFeO_3$ phase increase in an approximately linear fashion up to 870 $^oC$. Comparison of the normalised unit cell volume per $BiFeO_3$ unit is shown in Figure 4(a). It can be seen that this parameter increases with increasing temperature in the R3c phase to a value of 64.1 Å$^3$ before dropping sharply at $T_C$ to a value of approximately 63.1 Å$^3$ as a result of the phase change from R3c to Pbnm symmetry. Both the Fe-O and Bi-O bond lengths also undergo an abrupt change at $T_C$ (Figure 4). The octahedral bond environment in the R3c phase is comprised of 3 long degenerate Fe-O bond lengths and 3 short degenerate bond lengths. The longer of these distances increases initially to a maximum at around 500 $^oC$, consistent with the observations of Palewicz *et al.*[12], before decreasing as the phase transition temperature is approached, while the shorter Fe-O bond lengths slowly increase. In the Pbnm phase the $FeO_6$ octahedron is much more regular, with bond lengths being almost equal but occurring as three degenerate pairs, due to Fe being on an inversion centre. We note that there is no evidence for the antiferroelectric Fe displacements suggested by Haumont *et al.*[16]. The phase change from R3c to Pbnm also results in a change in the Bi-O environment. In the R3c phase the Bi is in a six co-ordinate site with respect to oxygen. This site is characterised by three degenerate long distances and three degenerate short distances. In both cases these increase with increasing temperature (unlike the Fe-O bond lengths). In contrast in the Pbnm phase the Bi is now in an 8 co-ordinate site with respect to oxygen, situated on a mirror plane (Table 1 and Figure 4). There is only one unique Bi-site, in contrast to the lower symmetry model suggested recently[16]. The Fe-O-Fe bond angle between neighbouring $FeO_6$ octahedra increases with increasing temperature on moving towards the α–β transition, consistent with previous neutron diffraction studies.[12] In the Pbnm phase there are two independent Fe-O-Fe bond angles; these increase only slightly over further temperature increases. Interestingly there is no significant change of the 'anti-phase' octahedral tilt on going from R3c ($a^-a^-a^-$) to Pbnm ($a^-a^-b^+$), with the value in the Pbnm phase at 830 °C remaining at about 10.8°. The corresponding in-phase tilt has a value of about 8.8°.

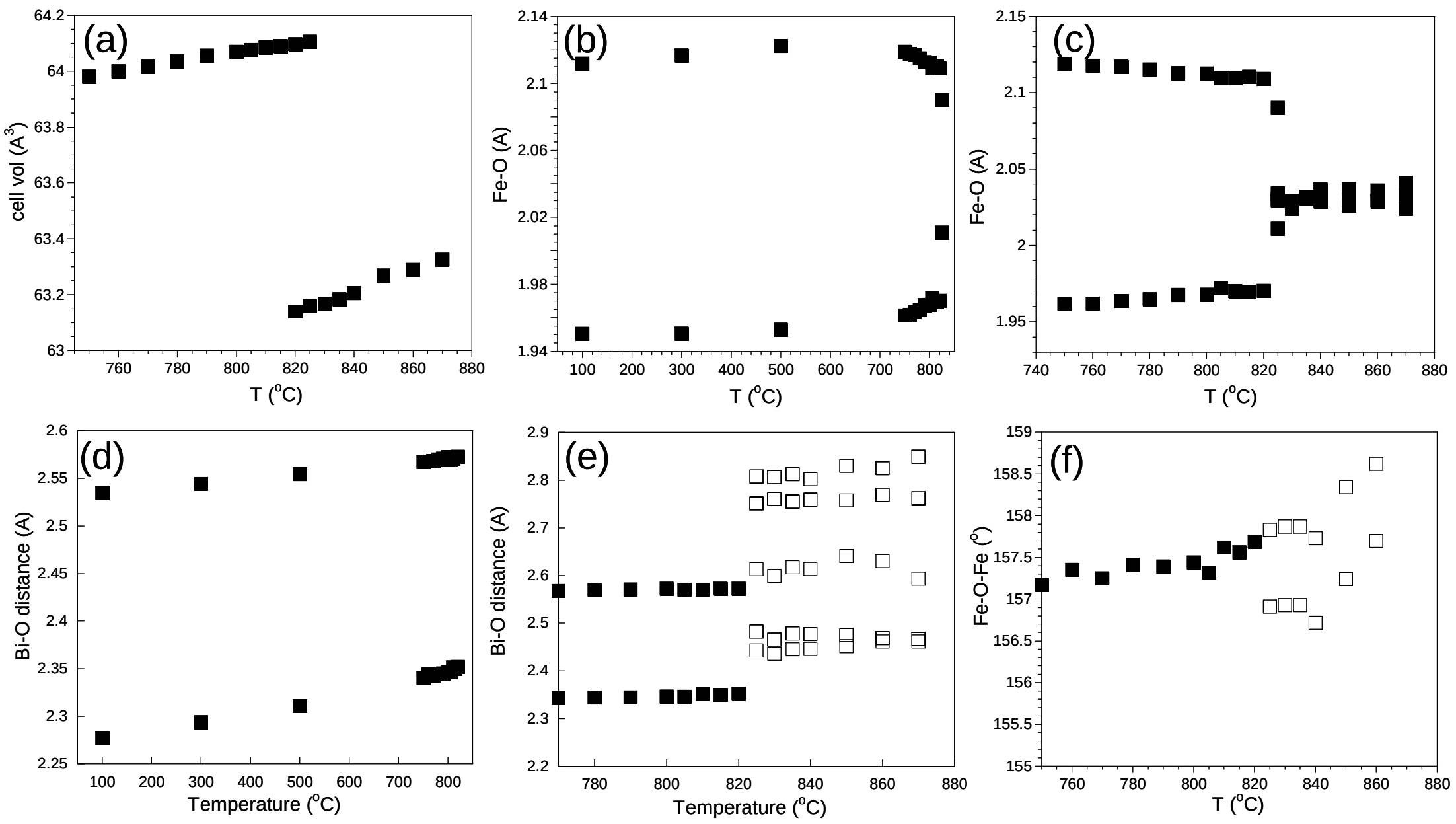

**Figure 4:** (a) Normalised unit cell volume per $BiFeO_3$ unit as a function of temperature showing an abrupt decrease at the α–β transition, (b) and (c) Evolution of Fe-O bond lengths, reflecting the loss of the polar displacement of $Fe^{3+}$ within its octahedron at $T_C$, (d) and (e) Evolution of the Bi-O environment showing the change from a 6 co-ordinate site in the R3c phase (closed squares) to a 8 co-ordinate site in the Pbnm phase (open squares) and (f) change in Fe-O-Fe bond angle in the R3c phase (closed squares) and the Pbnm phase (open squares).

The $Bi_2Fe_4O_9$ phase persists up to a temperature of 900 °C (the highest temperature investigated in this study) with no apparent phase transitions. The β-$BiFeO_3$ phase, however, disappears completely by 900 °C under the present experimental conditions. However, it is widely accepted that the rate of decomposition of $BiFeO_3$ into $Bi_2Fe_4O_9$ is not solely dependent on temperature but also on many other factors including exposure time at constant temperature, heating rate, ratio of surface to bulk volume, porosity and grain size.[13] This still suggests the enticing possibility of isolating the reported cubic γ-phase through careful control of experimental conditions.

In summary, we have demonstrated that $BiFeO_3$ undergoes a first order ferroelectric to paraelectric transition at $T_C$ (820 – 830 °C) from an α-phase of R3c symmetry to an orthorhombic β-phase. The latter phase is unambiguously identified as a $GdFeO_3$ –like perovskite, space group Pbnm, in contrast to all previous suggestions. This emphasises the need for PND rather than XRPD in the characterisation of such phase transitions. As far as we are aware there is no precedent for such a phase transition in a perovskite (ie. R3c to Pbnm): an *increase* in symmetry from orthorhombic to rhombohedral, with increasing temperature, as observed in $LaGaO_3$[29] is more usual. The nature of the transition is markedly different to the rhombohedral - orthorhombic transition in ferroelectric $BaTiO_3$, and to the rhombohedral – orthorhombic (ferroelectric – antiferroelectric) transition in $NaNbO_3$.[30] Further, careful experiments are necessary in order to shed further light on the existence and nature of a postulated γ-phase of $BiFeO_3$.

We thank the EPSRC, Royal Society and STFC for funding and Dr R. Goff and Ms A. Gibbs for experimental assistance.